\documentclass[prd,preprint,superscriptaddress,preprintnumbers,eqsecnum,showpacs,nofootinbib,nobibnotes]{revtex4}
\usepackage{amsfonts,amsmath,amssymb,bm,natbib}
\usepackage{graphicx}  

\newcommand{\be}{\begin{equation}}
\newcommand{\bea}{\begin{eqnarray}}
\newcommand{\ee}{\end{equation}}
\newcommand{\eea}{\end{eqnarray}}

\def\s#1{{\scriptscriptstyle #1}}

\def\noeq#1{(\ref{#1})}
\def\1eq#1{Eq.~(\ref{#1})}

\def\2eqs#1#2{Eqs.~(\ref{#1}) and~(\ref{#2})}
\def\3eqs#1#2#3{Eqs.~(\ref{#1}),~(\ref{#2}) and~(\ref{#3})}

\def\fig#1{Fig.~\ref{#1}}

\def\diff#1{{\rm d}^#1}

\def\ie{{\it i.e.}, }
\def\eg{{\it e.g.}, }

\def\n#1{({\it #1}\,)}

\def\nlsm{NL$\sigma$M }

\def\Tr{\mathrm{Tr}}
\def\G{\Gamma}
\def\lab#1{\lambda_{#1}^{\s{\rm B}}}
\def\la#1{\lambda_{#1}}

\def\rhob#1{\rho_{#1}^{\s{\rm B}}}

\def\vlam{{\cal V}^\lambda}
\def\vlamK{\vlam_{K_0}}
\def\vlamJ{\vlam_{\widetilde J}}

\def\Pi#1{{\cal P}^{({#1})}}
\def\PiK#1{\Pi{#1}_{K_0}}
\def\PiJ#1{\Pi{#1}_{\widetilde J}}
\def\Piphi#1{\Pi{#1}_{\phi}}

\def\cw{c_\s{W}}
\def\sw{s_\s{W}}

\begin{document}

\title{Renormalization Group Equation for \\
Weakly Power Counting Renormalizable Theories}

\author{D.~Bettinelli}
\email{bettinelli.daniele@gmail.com}
\affiliation{Dip. di Fisica, Universit\`a degli Studi di Milano,
via Celoria 16, I-20133 Milano, Italy}

\author{D. Binosi}
\email{binosi@ectstar.eu}
\affiliation{European Centre for Theoretical Studies in Nuclear
Physics and Related Areas (ECT*) and Fondazione Bruno Kessler, \\
Villa Tambosi, Strada delle
Tabarelle 286, 
I-38123 Villazzano (Trento)  Italy}

\author{A.~Quadri}
\email{andrea.quadri@mi.infn.it}
\affiliation{Dip. di Fisica, Universit\`a degli Studi di Milano,
via Celoria 16, I-20133 Milano, Italy}
\affiliation{INFN, Sezione di Milano, via Celoria 16, I-20133 Milano, Italy}

\begin{abstract}

We study the renormalization group flow in weak power counting (WPC) renormalizable theories. The latter are theories which, after being formulated in terms of certain variables, display only a finite number of independent divergent amplitudes order by order in the loop expansion. Using as a toolbox the well-known SU(2) non linear sigma model, we prove that for such theories a renormalization group equation holds that does not violate the WPC condition: that is, the sliding of the scale $\mu$ for physical amplitudes can be reabsorbed by a suitable set of finite counterterms arising at the loop order prescribed by the WPC itself. We explore in some detail the consequences of this result; in particular, we prove that it holds in the framework of a recently introduced beyond the Standard Model scenario in which one considers non-linear St\"uckelberg-like symmetry breaking contributions to the fermion and gauge boson mass generation mechanism.

\end{abstract}


\maketitle

\section{Introduction} 

Arguably, the most general way to look at the (perturbative) renormalization of quantum field theories is the one introduced two decades ago by Weinberg and Gomis~\cite{Gomis:1995jp}. In this modern look at the subject, the boundary between what constitutes a renormalizable or a non-renormalizable theory gets blurred. Indeed, in~\cite{Gomis:1995jp} it was proven that, if one includes in the tree-level vertex functional all possible Lorentz-invariant monomials in the fields, the external sources and their derivatives, while respecting, at the same time, the symmetries of the theory (encoded in the Batalin-Vilkovisky master equation~\cite{Gomis:1994he}), then it is possible to subtract all ultraviolet (UV) divergences by a redefinition of the tree-level parameters. What distinguishes between the renormalizable/non-renormalizable cases is the {\it stability} of the classical action under radiative corrections. In fact, in the renormalizable case the finite number of terms already present in the tree-level action is sufficient to absorb all UV divergences irrespectively of the loop order. This ceases to be true for non-renormalizable theories, where no matter how many terms are added to the tree-level action, there will always exist a sufficiently high order operator in the loop expansion which will give rise to a UV divergence that cannot be absorbed into the classical action. Such an action would therefore be {\it unstable} against radiative corrections, and one has to allow for {\it infinitely} many (symmetry preserving) terms in order to absorb all divergences.  

A prototype non-renormalizable theory (or, said in the positive, renormalizable in the modern sense), is the $d$-dimensional non-linear sigma model (\nlsm for short) in which the massless pseudoscalar ``pion'' fields $\phi^a$ form, together with the scalar field $\phi_0$ (with \mbox{$\langle\phi_0\rangle=m_d>0$}), a chiral multiplet subjected to the (nonlinear) constraint $\phi_a^2+\phi_0^2=m^2_d$; in this way the global symmetry \mbox{${\mathrm{SU}}(2)_\s{L}\times{\mathrm{SU}}(2)_{\s{R}}$} is non-linearly realized\footnote{In the 4-dimensional case $\phi_0$ is to be identified with the $\sigma$ meson and $m=m_4$ with the pion decay constant~$f_\pi$.} [see~\1eq{nltrans}].  Already at the one-loop level this theory shows an infinite number of divergent one-particle irreducible (1-PI) amplitudes, which in turn make the consistent subtraction of UV divergences problematic. However, by embedding the global \nlsm  into a locally symmetric formulation in terms of a classical source corresponding to a certain (flat) connection $\widetilde J^a_\mu$, it was shown in~\cite{Ferrari:2005ii} that one can acquire full control over the UV divergences of the model. Specifically, it can be shown that in such a theory a Weak Power Counting (WPC) exists~\cite{Ferrari:2005va,Quadri:2014gza} which implies that  the number of independent divergent amplitudes stays finite at each loop order $n$, even though it increases with $n$. These ``ancestor'' amplitudes, are not the ones associated with the quantum pion fields $\phi^a$, but are rather written in terms of the connection $\widetilde J^a_\mu$ and the composite operator $K_0$ implementing the aforementioned non-linear constraint.  Then, the presence of the so-called Local Functional Equation (LFE) together with a suitable change of field variables called ``bleaching'', allows one to generate from the ancestor amplitudes all the (infinite) descendant (off-shell) amplitudes involving the pion fields, uniquely fixing {\it en route} their divergent part at any order in the loop expansion~\cite{Quadri:2014gza}.

Within the aforementioned Weinberg-Gomis approach to renormalization, one can interprete the WPC as a condition dictating at which loop order the coefficient of a particular monomial in the ancestor variables and their derivatives becomes non-vanishing. While it has been proven in~\cite{Bettinelli:2007zn} that the WPC allows for the definition of a symmetric (\ie compatible with the symmetries of the theory) subtraction scheme, the question has remained open of whether a Renormalization Group (RG) equation compatible with the WPC exists. The problem can be stated in the following terms. WPC renormalizable theories possess a RG flow associated to the RG equation for ancestor amplitudes. Imagine then that a change in the scale $\mu$ of the radiative corrections is inequivalent to a rescaling of the coefficients of the ancestor amplitudes counterterms, at the given loop order prescribed by the WPC: this would imply that the RG flow mixes up the hierarchy of UV divergences encoded in the WPC, thus making it impossible to slide the scale $\mu$ between different energies.   

As we will prove in this paper, fortunately this is not the case, as indeed {\it the RG equation of ancestor amplitudes turns out to be compatible with the WPC}. This is an important result which acquires particular relevance in the context of the non-linearly realized electroweak theory introduced in~\cite{Binosi:2012cz,Bettinelli:2013hia}, in which the WPC has been used as a model building principle. In particular, in this model the classical source $\widetilde J^a_\mu$ is promoted to a local dynamical field responsible for generating part of the mass of the $W^\pm$ and $Z$ gauge bosons through the St\"uckelberg mechanism; this leads in turn to many unique features which cannot be found in theories describing physics beyond the Standard Model (\eg it is impossible to add a scalar singlet without breaking the WPC, so that the minimal number of physical scalar resonances in the model is 4). As we will see, the result proven here for the compatibility between the RG flow and the WPC applies also in this case. This entails the possibility of evolving the scale $\mu$ in a mathematically consistent way, thus allowing to obtain  predictions for the relevant observables applicable in different energy regimes and thus paving the way for a systematic study of their deviations from the expected SM results.

The paper is organized as follows. By using the aforementioned example of the $d$-dimensional SU(2) \nlsm in Sect.~\ref{prel} we review the embedding of the model in a local formulation in terms of a flat connection, leading to the LFE and the WPC condition. The concepts of ancestor and descendant amplitudes as well as of bleached variables are introduced together with the corresponding symmetric scheme for subtracting the UV divergences. In Sect.~\ref{renorm} we derive the equation governing the RG flow in the local \nlsm. Next, after specializing this equation to ancestor amplitudes, we discuss under which conditions the RG flow preserves the WPC; we then prove that at the one-loop level these conditions are indeed satisfied. Sect.~\ref{genth} contains the central result of the paper: there we prove the general theorem stating that the RG-flow of ancestor amplitudes preserves the WPC. In Sect.~\ref{stability} we introduce the notion of weak stability, while in Sect.~\ref{bsm} we derive the consequences of the RG flow theorem for the aforementioned non-linearly realized SU(2)$\times$U(1) electroweak theory. Our conclusions and outlook are finally presented in Sect.~\ref{concl}.

\section{\label{prel}Preliminaries}


\subsection{Global \nlsm}

The $d$-dimensional action of a \nlsm is conventionally written as
\begin{equation}
S_0=\frac{m^2_d}{4g^2}\int\!\diff{d}x\,\Tr\left[\partial_\mu\Omega^\dagger\partial^\mu\Omega
\right],
\label{nlsm-cl-act}
\end{equation}
where $\Omega=\Omega(x)$ (with $x$ the space-time coordinates) represents a matrix belonging to a symmetry group $G$.
$m_d=m^{d/2-1}$ is the mass scale of the theory ($m$ has dimension 1).
We will consider in particular the case where $G$ is $SU(N)$ (and then
specialize to the case $N=2$):
\begin{align}
\Omega^\dagger\Omega=1;\qquad {\mathrm{det}}\,\Omega=1 \, .
\end{align}

In terms of a suitable basis of fields $\phi^a(x)$, parametrizing the $G$ matrix $\Omega$, \ie \mbox{$\Omega(x)=\Omega(\phi^a(x))$}, the action~\noeq{nlsm-cl-act} reads
\begin{align}
S_0&=\frac{m^2_d}2\int\!\diff{d}x\,g_{ab}\partial_\mu\phi^a\partial^\mu\phi^b;&
g_{ab}=g_{ba}=\frac1{2g^2}\Tr\left[\frac{\partial\Omega}{\partial\phi^a}\frac{\partial\Omega^\dagger}{\partial\phi^b}\right].
\label{nlsm-cl-act1}
\end{align}
The fields $\phi_a$ will be generically referred to as ``pion'' fields. 
Geometrically they represent the coordinates of the group manifold $G$.

In what follows we will mostly deal with the case $G\equiv{\mathrm{SU}}(2)$, where one can set
\begin{align}
\Omega&=\frac1{m_d}\left(\phi_0+ig\phi_a\tau_a\right);& &\phi_0^2+g^2\phi_a^2=m^2_d;&
\phi_0&=\sqrt{m^2_d-g^2\phi_a^2},
\label{nlc}
\end{align}
where $a=1,2,3$, and $\tau_a$ are the usual Pauli matrices.
The action~\noeq{nlsm-cl-act1} will then contain non-polynomial derivative interactions involving the pion fields, reading
\begin{equation}
S_0=\int\!\diff{d}x\left[\frac12\partial_\mu\phi_a\partial^\mu\phi_a+\frac12g^2\frac{(\phi_a\partial_\mu\phi_a)(\phi_b\partial_\mu\phi_b)}{\phi_0^2}
\right].
\end{equation}
It is exactly the presence of two derivatives in the interaction term that in $d>2$ causes severe UV divergences, which are ultimately responsible for the non-renormalizability of the corresponding quantized theory.

When writing the action as in~\noeq{nlsm-cl-act}, it is immediate to show that the theory is invariant under a non-linearly realized ${\mathrm{SU}}(2)_\s{L}\times{\mathrm{SU}}(2)_{\s{R}}$ global symmetry:
\begin{align}
\Omega\to U\Omega V^\dagger;\qquad U\in{\mathrm{SU}}(2)_\s{L}, \quad V\in{\mathrm{SU}}(2)_\s{R}.
\end{align}
In terms of the pion fields, the left infinitesimal transformation of constant parameters $\omega_a$ reads
\begin{equation}
\delta\phi_0(x)=-\frac12g^2\omega_a\phi_a(x);\qquad
\delta\phi_a(x)=\frac12\omega_a\phi_0(x)+\frac12g\epsilon_{abc}\phi_b(x)\omega_c.
\label{nltrans}
\end{equation}
In matrix form the infinitesimal left transformation is
\begin{equation}
\delta \Omega =ig \omega_a \frac{\tau_a}{2} \Omega \, .
\label{nl.matrix}
\end{equation}

\subsection{Local \nlsm}

For any unitary matrix $\Omega$ it is possible to define a flat connection\footnote{The term ``flat'' refers to the fact that the field strength associated to $F_\mu$ vanishes.}
\begin{equation}
F_\mu=\frac ig\Omega\partial_\mu\Omega^\dagger,\end{equation}
so that the action~(\ref{nlsm-cl-act}) can be cast in the form
\begin{equation}
S_0=\frac{m^2_d}{4}\int\!\diff{d}x\,\Tr\left[F_\mu F^\mu\right].
\label{nlsm-cl-act2}
\end{equation}

Specialize now to the SU(2) case, and consider a local SU(2)$_\s{L}$ transformation on $\Omega$; this will induce a gauge transformation on the flat connection $F_\mu$, namely
\begin{equation}
\Omega\to U\Omega\quad\Longrightarrow\quad
F_\mu\to UF_\mu U^\dagger+\frac ig U\partial_\mu U^\dagger.
\label{lgt1}
\end{equation}
Clearly, the action~\noeq{nlsm-cl-act2} is not invariant under these local transformations. 

However, let us introduce an additional classical source $\widetilde J_\mu$ transforming as a gauge connection under the local SU(2)$_\s{L}$ group. At this point the difference $F_\mu -\widetilde J_\mu$ will transform in the adjoint representation
\begin{equation}
I_\mu=F_\mu-\widetilde J_\mu\to UI_\mu U^\dagger,
\end{equation}
so that the action
\begin{equation}
S=\frac{m^2_d}{4}\int\!\diff{d}x\,\Tr\left[I_\mu I^\mu\right]
\label{lgt2}
\end{equation}
is invariant under a local SU(2)$_\s{L}$ symmetry. In coordinates one has
\begin{equation}
X_\mu=\frac12 X^a_\mu\tau_a;\qquad X=F,\widetilde{J}, I,
\end{equation}
with
\begin{align}
F^a_\mu&=\frac2{m^2_d}\left[\phi_0\partial_\mu\phi^a-\phi^a\partial_\mu\phi_0+g\epsilon^{abc}(\partial_\mu\phi^b)\phi^c\right];& I^a_\mu=F^a_\mu-\widetilde J^a_\mu \, .
\label{exp}
\end{align}
The local infinitesimal transformations are
\begin{align}
\delta\phi_0(x)&=-\frac12g^2\omega_a(x)\phi_a(x);&
\delta\phi_a(x)&=\frac12\omega_a(x)\phi_0(x)+\frac12g\epsilon_{abc}\phi_b(x)\omega_c(x);\nonumber \\
\delta \widetilde J^a_\mu(x)&=\partial_\mu\omega_a(x)+g\epsilon_{abc} \widetilde J^b_\mu(x)\omega_c(x).
\label{trans}
\end{align}

The global \nlsm is embedded in the local formulation we have just provided. Specifically,  the terms $\widetilde JF$ and $\widetilde J^2$ are separately 
invariant under a global SU(2)$_\s{L}$ transformation
(that is when the $\omega$ gauge parameters are kept constants); therefore we can set $\widetilde J$ directly to zero to obtain
\begin{equation}
S_0=\left. S\right\vert_{\widetilde J=0}.
\end{equation}

\subsection{Local Functional Equation}

The advantage of the gauged formulation of the \nlsm provided by the action~\noeq{lgt2} resides in the existence of a functional identity that can be obtained by exploiting the invariance of the Haar path integral measure under the local gauge transformations~\noeq{trans}. This equation, which goes under the name of Local Functional Equation (LFE) reads~\cite{Ferrari:2005ii}
\begin{equation}
-\partial_\mu \frac{\delta \G^{(0)}}{\delta \widetilde J^a_{\mu}(x)} + 
g\epsilon_{abc} \widetilde J^c_{\mu}(x) \frac{\delta \G^{(0)}}{\delta \widetilde J^b_{\mu}(x)}+
\frac{1}{2} \frac{\delta \G^{(0)}}{\delta K_0(x)} \frac{\delta \G^{(0)}}{\delta \phi_a(x)} + \frac{1}{2}g \epsilon_{abc} \phi_c(x) 
\frac{\delta \G^{(0)}}{\delta \phi_b(x)}= -\frac{1}{2}g^2 \phi_a(x) K_0(x), 
\label{LFEcl}
\end{equation}
where $\Gamma^{(0)}$ is given by
\begin{equation}
\Gamma^{(0)}=S+S_{\rm ext};\qquad S_{\rm ext}=\int\!\diff{d}x\, K_0\phi_0,
\label{tree-lev}
\end{equation}
$K_0$ being an SU(2)$_\s{L}$ invariant source associated to the non-linear constraint~\noeq{nlc}. Thus, $K_0$ is associated with the auxiliary external field required to define the composite operator entering in the non-linear symmetry transformation. It plays the same role as the antifields~\cite{Gomis:1994he} in gauge theories and the tree-level dependence of the vertex functional on $K_0$ is fixed by the form of the non-linear transformation~\noeq{trans}: since the only composite operator entering in $\delta \phi_a$ is $\phi_0$, one only
needs an external source $K_0$~\cite{Quadri:2014gza}.

As the symmetry is non-anomalous, the LFE~\noeq{LFEcl} is satisfied by the full vertex functional~$\G$:
\begin{equation}
-\partial_\mu \frac{\delta \G}{\delta \widetilde J^a_{\mu}(x)} + 
g\epsilon_{abc} \widetilde J^c_{\mu}(x) \frac{\delta \G}{\delta \widetilde J^b_{\mu}(x)}+
\frac{1}{2} \frac{\delta \G}{\delta K_0(x)} \frac{\delta \G}{\delta \phi_a(x)} + \frac{1}{2}g \epsilon_{abc} \phi_c(x) 
\frac{\delta \G}{\delta \phi_b(x)}= -\frac{1}{2}g^2 \phi_a(x) K_0(x) \, .
\label{LFE}
\end{equation}
In addition, notice that since the rhs of~\1eq{LFE} 
is linear in the quantized fields and thus 
remains classical, this term will be present only at tree-level. 

The LFE encodes at the quantum level the classical local SU(2) transformation, whose form in Eq.(\ref{trans}) is not preserved under radiative corrections. 
Therefore one cannot constrain the quantum 1-PI Green's functions 
on the basis of the classical symmetry in Eq.~(\ref{trans}); the constraints
satisfied by these functions are fixed by Eq.(\ref{LFE}).

One of its main consequences is the separation of the 1-PI amplitudes into two classes. On the one hand, there are the amplitudes involving only the insertion of the SU(2) connection $\widetilde J^a_\mu$ and of the source of the non-linear constraint $K_0$: these are called ancestor amplitudes~\cite{Ferrari:2005ii,Ferrari:2005va}. On the other hand, we have the so-called descendant amplitudes, \ie those involving at least one external $\phi$-leg. These amplitudes are not independent, as they are uniquely determined by the LFE once the ancestor amplitudes are known~\cite{Ferrari:2005ii,Ferrari:2005va,Bettinelli:2007kc}.

\subsection{Weak Power Counting}

Despite the fact that they do not involve external legs of the quantized fields of the theory $\phi_a$, the truly fundamental Green's functions of the local \nlsm are the ancestor amplitudes. Such Green's functions display an UV behaviour that is significantly better than the one of their descendants: namely, there exists a choice of the tree-level action, compatible with the symmetries of the theory, such that only a finite number of divergent ancestor amplitudes arises order by order in the loop expansion.
This property is dubbed the Weak Power-Counting (WPC) condition~\cite{Ferrari:2005va}.

Indeed, one can show that in $d$-dimensions an $n$-loop ancestor amplitude $G$ with $N_{\widetilde J}$ ($N_{K_0}$) external $\widetilde J_\mu^a$ ($K_0$) legs has a superficial degree of divergence given by~\cite{Ferrari:2005va}
\begin{equation}
D(G) = (d-2)n + 2 - N_{\tilde J} - 2 N_{K_0}.
\label{WPC}
\end{equation}
Thus, at every-loop order only a finite number of superficially divergent ancestor amplitudes exists, \ie the ones for which $D(G) \geq 0$; obviously, the local \nlsm is still non renormalizable, as~\1eq{WPC} shows that as $n$ grows bigger the number of UV divergent amplitudes increases. For example, in the 4-dimensional 
case~\1eq{WPC} tells us that at one-loop the UV divergent amplitudes involve up to four external $\widetilde J^a_\mu$ legs and/or two $K_0$ legs. Accordingly, the one-loop 1PI functional for these ancestor amplitudes reads
\begin{align}
{\cal A}^{(1)}[K_0,\widetilde{J}^a_\mu]&=\frac12\int\!\G^{(1)}_{\widetilde J^a_\mu\widetilde J^b_\nu}(x,y)\widetilde J^a_\mu(x)\widetilde J^b_\nu(y)+\frac1{3!}\int\!\G^{(1)}_{\widetilde J^a_\mu\widetilde J^b_\nu\widetilde J^c_\rho}(x,y,z)\widetilde J^a_\mu(x)\widetilde J^b_\nu(y)\widetilde J^c_\rho(z)\nonumber \\
&+\frac1{4!}\int\!\G^{(1)}_{\widetilde J^a_\mu\widetilde J^b_\nu\widetilde J^c_\rho \widetilde J^d_\sigma}(x,y,z,w)\widetilde J^a_\mu(x)\widetilde J^b_\nu(y)\widetilde J^c_\rho(z)\widetilde J^d_\sigma(w)+\frac12\int\!\G^{(1)}_{K_0 K_0}(x,y)K_0(x)K_0(y)\nonumber \\
&+\frac12\int\!\G^{(1)}_{K_0\widetilde J^a_\mu\widetilde J^b_\nu}(x,y,z)K_0(x)\widetilde J^a_\mu(y)\widetilde J^b_\nu(z) + \cdots
\label{ancestor1PI}
\end{align}
where the dots stand for ancestor amplitudes that are not UV-divergent at one loop.

\subsection{Bleached variables}

Ancestor amplitudes {\it per-se} are not a solution of the LFE~\noeq{LFE} as they carry no information for amplitudes involving pion fields. To achieve this, it is necessary to introduce invariant combinations in one-to-one correspondence to the ancestor variables $\widetilde J^a_\mu$ and $K_0$. These so-called bleached variables are found to be~\cite{Ferrari:2005va}
\begin{align}
j_\mu&=\Omega^\dagger I_\mu\Omega=\frac12j^a_\mu\tau^a;&
\overline{K}_0 &= \frac{m_d^2 K_0}{\phi_0} - 
\phi^a \frac{\delta S}{\delta \phi^a}.\nonumber \\
&\left.-j_\mu^a\right\vert_{\phi^a=0}=\widetilde{J}^a_\mu;&
&\left.\overline{K}_0\right\vert_{\phi^a=0}=m_d K_0,
\label{bleached}
\end{align}
where the action $S$ appearing in the definition of $\overline{K}_0$ is given in~\1eq{lgt2}. 

In terms of bleached variables, the one-loop version of the LFE can be cast in the form~~\cite{Bettinelli:2007kc}
\begin{equation}
\frac{\partial}{\partial\phi_b}\Gamma^{(1)}[\phi^a,\overline{K}_0,j^a_\mu]=0,
\end{equation}
and one can prove that the complete solution of the  one-loop LFE is given by
\begin{equation}
\Gamma^{(1)}[\phi^a,\overline{K}_0,j^a_\mu]=\left.{\cal A}^{(1)}[K_0,\widetilde{J}^a_\mu]\right\vert^{
K_0\to\overline{K}_0/m_d}_{\widetilde{J}^a_\mu\to -j^a_\mu},
\end{equation} 
where ${\cal A}^{(1)}$ is the 1PI functional of the ancestor amplitudes~\noeq{ancestor1PI}. Thus one finds that at this order, all the dependence on the pion fields is enclosed in the bleached variables.

This prescription allows one to write down all the descendant amplitudes depending on the pion fields in terms of the ancestor amplitudes; all one has to do is to expand the bleached variables up to the relevant order in the pion fields. A direct computation using the definition~\noeq{bleached} shows that $\overline K_0$ starts with two $\phi$'s, with
%
%
%
\begin{align}
\overline K_0=m_d K_0-\frac{m_d}{2}\phi^a\partial_\mu\widetilde J^\mu_a-g\epsilon^{abc}\phi^a(\partial_\mu\phi^b)\widetilde J^\mu_c
+\phi^a\square\phi^a+ \frac{g^2}{2 m_D} K_0 \phi_a^2 + \cdots.
\label{barK0exp}
\end{align}
For the variable $j^a_\mu$ one has instead the result
\begin{align}
m_d^2 j^a_{\mu} & =  m_d^2 I^a_{\mu} - 2 g^2\phi_b^2 I^a_{\mu} + 2 g^2\phi_b I^b_{\mu} \phi_a +2g \phi_0 \epsilon_{abc} \phi_b I^c_{\mu},
\label{e.7}
\end{align}
yielding, with the help of~\1eq{exp}, to the expansion
\begin{align}
j_\mu^a&=-\widetilde{J}^a_\mu+\frac2{m_d}\partial_\mu\phi^a-\frac2{m_d^2}g\epsilon_{abc}(\partial_\mu\phi^b)\phi^c+\frac2{m^2_d}g\left(g\phi^2_c\delta^{ab}-g\phi^a\phi^b+m_d\epsilon^{abc}\phi^c\right)\widetilde J^b_\mu
+\cdots.
\label{jexp}
\end{align}
Thus, for example, one has for the two- and three-point pion sector
\begin{align}
\frac12\int\!\G^{(1)}_{\phi^a\phi^b}(x,y)\phi^a(x)\phi^b(y)
&=\frac2{m^2_d}\int\!\G^{(1)}_{\widetilde J^a_\mu\widetilde J^b_\nu}(x,y)\partial^\mu\phi^a(x)\partial^\nu\phi^b(y),\nonumber \\
\frac1{3!}\int\!\G^{(1)}_{\phi^a\phi^b\phi^c}(x,y,z)\phi^a(x)\phi^b(y)\phi^c(z)&=-\frac4{m^3_d}g\int\!\G^{(1)}_{\widetilde J^a_\mu\widetilde J^b_\nu}(x,y)\epsilon^{acd}[\partial^\mu\phi^c(x)]\phi^d(x)\partial^\nu\phi^b(y)\nonumber \\
&+\frac4{3m^3_d}\int\!\G^{(1)}_{\widetilde J^a_\mu\widetilde J^b_\nu\widetilde J^c_\rho}(x,y,z)[\partial^\mu\phi^a(x)][\partial^\nu\phi^b(y)][\partial^\rho\phi^c(z)].
\end{align}

At higher orders in the loop expansion the bilinearity of the LFE
(\ref{LFE}) implies that there is an explicit dependence on the pion fields, governed by the equation~\cite{Bettinelli:2007kc}
\begin{equation}
\frac{\delta\G^{(n)}}{\delta\phi^a(x)}=-\frac12\sum_{i=1}^{n-1}\frac{\delta\G^{(i)}}{\delta\overline{K}_0(x)}\frac{\delta\G^{(n-i)}}{\delta\phi^a(x)}.
\label{bilieq}
\end{equation}

Notice that the above form of the LFE holds provided that $\G$ is written as a functional of the variables $\overline{K}_0, j_{a\mu}$.

The general solution of the LFE becomes then~\cite{Bettinelli:2007kc}
\begin{equation}
\Gamma[\phi^a,\overline{K}_0,j^a_\mu]=\left.{\cal A}[K_0,\widetilde{J}^a_\mu]\right\vert^{
K_0\to\overline{K}_0/m_d}_{\widetilde{J}^a_\mu\to -j^a_\mu}+{\cal G}[\phi^a,\overline{K}_0,j^a_\mu],
\label{fullLFEsol}
\end{equation}
where ${\cal G}$ is the functional solving~\1eq{bilieq}; as such it is uniquely fixed by the ancestor amplitudes,  depends explicitly on $\phi_a$, and, finally, vanishes at $\phi_a = 0$. The existence of ${\cal G}$ can be proven by exploiting cohomological tools~\cite{Bettinelli:2007kc}.

\subsection{Renormalization}

Summarizing, the combination of the LFE and the WPC, expressed in~\2eqs{LFE}{WPC} respectively, allows one to express the infinite number of divergent amplitudes involving the pion fields (descendant amplitudes) in terms of a finite number of ancestor amplitudes involving the connection $\widetilde J^a_\mu$ and the source of the non-linear constraint~\noeq{nlc} $K_0$.    

It turns out that it is also possible to renormalize the theory in a symmetric fashion, that is, in a way that preserves the LFE~\cite{Bettinelli:2007zn,Ferrari:2005va}. 

Consider first the one-loop ancestor amplitudes. Taking into account Lorentz and global SU(2)${_\s R}$ invariance, the list of UV divergent amplitudes reduces to the following eight (integrated) local monomials
\begin{align}
{\cal M}_0&=\int\!\diff{d}x\,(\widetilde J^a_\mu\widetilde J_a^\mu);&
{\cal M}_1&=\int\!\diff{d}x\,(\partial_\mu\widetilde J_a^\mu)(\partial_\nu\widetilde J_a^\nu);&
{\cal M}_2&=\int\!\diff{d}x\,(\partial_\mu\widetilde J^a_\nu)(\partial^\mu\widetilde J_a^\nu);
\nonumber \\
{\cal M}_3&=\int\!\diff{d}x\,\epsilon_{abc}(\partial_\mu\widetilde {J}^a_\nu)\widetilde J_b^\mu \widetilde J_c^\nu;&
{\cal M}_4&=\int\!\diff{d}x\,(\widetilde J^a_\mu\widetilde J_a^\mu)(\widetilde J^b_\nu\widetilde J_b^\nu);& 
{\cal M}_5&=\int\!\diff{d}x\,(\widetilde J^a_\mu\widetilde J_b^\mu)(\widetilde J^a_\nu\widetilde J_b^\nu);
\nonumber \\
{\cal M}_6&=\int\!\diff{d}x\,(K_0)^2;&
{\cal M}_7&=\int\!\diff{d}x\,K_0(\widetilde J^a_\mu\widetilde J_a^\mu).&
\label{theMs}
\end{align}
For example, for the one-loop two-point function one finds~\cite{Ferrari:2005va}
\begin{equation}
\G^{(1)}_{\widetilde J\widetilde J}
=\left(-\frac1{12}\frac1{d-4}\frac{m^2_d}{m^2}\frac{g^2}{(4\pi)^2}+\cdots\right)\int\!\diff{d}x\,\widetilde J^\mu_a(\square g_{\mu\nu}-\partial_\mu\partial_\nu)\widetilde J^\nu_a,
\end{equation}
where $m\equiv m_{d=4}$, and the dots indicate finite ($\mu$-dependent) pieces. Then, we can dispose of this divergence by requiring that the monomials ${\cal M}_{1,2}$ enters the counterterm action in the combination
\begin{equation}
\rho_1^\s{\rm B}{\cal M}_1+\rho_2^\s{\rm B}{\cal M}_2=\left(\frac1{12}\frac1{d-4}\frac {m_d^2}{m^2}\frac{g^2}{(4\pi)^2}+\cdots\right)({\cal M}_1-{\cal M}_2).
\end{equation}
Carrying out this procedure for all of the one-loop divergent ancestor amplitudes, one can fix the one-loop counterterm action $S_{\rm ct}=\sum_i\rho_i^\s{\rm B}\,{\cal M}_i$ thus rendering finite the \nlsm at this level in the loop expansion.

The one-loop counterterms for the descendant amplitudes are then generated by expressing the  monomials ${\cal M}_i$ appearing in $S_{\rm ct}$ in terms of the bleached variables, whence giving rise to the SU(2)$_\s{L}$ invariants
\begin{align}
{\cal M}_0\to&\hspace{0.2cm}{\cal I}_0=\int\!\diff{d}x\,(j^a_\mu j_a^\mu)=\int\!\diff{d}x\,(I^a_\mu I_a^\mu),
\nonumber \\
{\cal M}_1\to&\hspace{0.2cm}{\cal I}_1=\int\!\diff{d}x\,(\partial_\mu j_a^\mu)(\partial_\nu j_a^\nu)=\int\!\diff{d}x\,({\cal D}^{ab}_\mu I^\mu_b)({\cal D}^{ac}_\nu I^\nu_c),\nonumber \\
{\cal M}_2\to&\hspace{0.2cm}{\cal I}_2=\int\!\diff{d}x\,(\partial_\mu j^a_\nu)(\partial^\mu j_a^\nu)=\int\!\diff{d}x\,({\cal D}^{ab}_\mu I^b_\nu)({\cal D}_{ac}^\mu I_c^\nu),\nonumber \\
-{\cal M}_3\to&\hspace{0.2cm}{\cal I}_3=\int\!\diff{d}x\,\epsilon_{abc}(\partial_\mu j^a_\nu) j_b^\mu j_c^\nu=\int\!\diff{d}x\,\epsilon_{abc}({\cal D}^{ad}_\mu I^d_\nu) I_b^\mu I_c^\nu,\nonumber \\
{\cal M}_4\to&\hspace{0.2cm}{\cal I}_4=\int\!\diff{d}x\,(j^a_\mu j_a^\mu)(j^b_\nu j_b^\nu)=\int\!\diff{d}x\,(I^a_\mu I_a^\mu)(I^b_\nu I_b^\nu),\nonumber\\ 
{\cal M}_5\to&\hspace{0.2cm}{\cal I}_5=\int\!\diff{d}x\,(j^a_\mu j_b^\mu)(j^a_\nu j_b^\nu)=\int\!\diff{d}x\,(I^a_\mu I_b^\mu)(I^a_\nu I_b^\nu),\nonumber \\
m^2_d{\cal M}_6\to&\hspace{0.2cm}{\cal I}_6=\int\!\diff{d}x\,(\overline K_0)^2,\nonumber \\
m_d{\cal M}_7\to&\hspace{0.2cm}{\cal I}_7=\int\!\diff{d}x\,\overline K_0(j^a_\mu j_a^\mu)=\int\!\diff{d}x\,\overline K_0(I^a_\mu I_a^\mu),
\label{theIs}
\end{align}
in which all the covariant derivatives are defined with respect to the flat connection:
\begin{equation}
{\cal D}^{ac}_\mu=\partial_\mu\delta^{ac}+g\epsilon^{abc}F^b_\mu.
\end{equation}

It turns out that there is no one-loop counterterm associated with ${\cal I}_0$,
as the theory is massless~\cite{Ferrari:2005va}.
Since at $\phi^a=0$ one has the normalization conditions given in
the second line of~\1eq{bleached}, clearly we recover the counterterms introduced  for the one-loop ancestor amplitudes; in addition, however, the above invariants generate the correct one-loop counterterms for all pion amplitudes, solving completely the hierarchy imposed by the LFE at this order. 

At higher orders, say $n > 1$, the situation is slightly more complicated. The bilinear term in the LFE results in the term~\noeq{bilieq}; however, this can only give rise to the mixing of lower order counterterms and therefore does not lead to new ones. As a consequence, this term will not appear in the evaluation of the $n^{\rm th}$-order counterterms for the ancestor amplitudes. 
Thus the symmetric subtraction procedure at order $n$ is the following. One starts by computing the divergent part of the ancestor amplitudes that are superficially divergent according to the WPC condition~\noeq{WPC}. This will then fix the coefficients of the local monomials ${\cal M}_i$ appearing at this order. Then one converts these monomials into the invariants ${\cal I}_i$ by writing them in terms of the bleached variables. This will then give rise to the full set of counterterms required to make the theory finite at order $n$ in the loop expansion.

\section{\label{renorm}A renormalization group equation for WPC renormalizable theories}

From the discussion of the previous section we know that the LFE holds true for the effective action $\widehat\G$ of the theory, which comprises the tree-level Feynman rules plus counterterms.
In addition, in the case of zero pion fields,
the full bare effective action $\widehat\Gamma^\s{\rm B}_0\equiv\left.
\widehat\G^\s{\rm B}\right\vert_{\phi^a=0}$ can be decomposed on a basis of integrated local monomials involving only the variables $K_0$ and $\widetilde J^a_\mu$ and their 
derivatives
\begin{equation}
\widehat\Gamma_0^\s{\rm B}=
\sum_i \rhob{i} \, {\cal M}_i(K_0,\widetilde J^a_{\mu}).
\label{e.11}
\end{equation}
In the above equation the sum spans all possible (infinite) local monomials, compatible with Lorentz invariance. By expressing the bare parameters $\rhob{i}$ in terms of the renormalized ones $\rho_i$ and of the scale $\mu$ of the radiative corrections, one gets the effective action $\widehat\G$, which yields a finite theory.
Thus, from the point of view of Weinberg and Gomis renormalizability, the WPC selects which coefficients $\rhob{i}$ must be zero in the tree-level approximation, and prescribes the loop order at which the counterterms of a given local monomial starts to appear (or, in other words, at which order a particular coefficient $\rhob{i} \neq 0$). 

Knowledge of the rhs of~\1eq{e.11} completely fixes (through the LFE) the dependence on the pion fields of the complete bare action $\widehat\G^\s{\rm B}$. Then, one can reabsorb the dependence on the scale $\mu$ of the radiative corrections into the renormalized parameters $\rho_i$ by expressing the bare parameters in terms of the renormalized ones:
\begin{equation}
\widehat\G[\rhob{i}]=\widehat\G[\rho_i,\mu].
\label{RGR.ct}
\end{equation}
This is always possible, since due to the linearity of
$\widehat\Gamma_0^\s{\rm B}$ on the bare parameters, one can reabsorb
the divergences associated with the monomial ${\cal M}_i$ by
redefining the bare parameter~$\rhob{i}$. In addition,~\1eq{RGR.ct} entails that the same result holds for the full vertex functional $\G$, namely
\begin{equation}
\G[\rhob{i}]=\G[\rho_i,\mu],
\label{RGR}
\end{equation}
Next, by differentiating~\1eq{RGR} with respect to the scale $\mu$, we get the following RG equation
\begin{equation}
\mu \frac{\partial \G}{\partial \mu} + \sum_i \mu
\frac{\partial \rho_{i}}{\partial \mu} \frac{\partial \G}{\partial \rho_{i}} =0,
\label{RGE}
\end{equation}
which holds in full generality for all 1-PI Green's functions (including those involving an explicit dependence on the pion legs)\footnote{In the Chiral Lagrangian approach (momentum expansion) a RG equation has been derived in~\cite{Buchler:2003vw}.}.

The very important question we are addressing in this paper is whether the WPC condition is compatible with the RG flow controlled by the RG-equation for the ancestor amplitudes. In general, in fact, it might happen that a change in the scale $\mu$ of the radiative corrections is not equivalent to a rescaling of the coefficients of the ancestor amplitudes counterterms, at the given loop order prescribed by the WPC. In that case the RG flow would mix up the hierarchy of UV divergences encoded in the WPC, thus making it impossible to slide the scale $\mu$ between different energies.

Let us start addressing this question by noticing that, for the zero (external) pion fields case, \1eq{RGE} gives rise to a particularly simple relation. Indeed in this case $\widehat\G_0$ is linear\footnote{This is definitely not the case for the complete effective action $\widehat\G$, as in this case the functional ${\cal G}$ appearing in the complete LFE solution~\noeq{fullLFEsol} contains a product of lower order terms and therefore it has a complicated dependence on the $\rho_i$.} in the bare parameters $\rhob{i}$, 
and therefore a change in the scale $\mu$, affecting the $n^{\rm th}$ order action, can be accommodated by a change of the finite part of the $n^{\rm th}$  order (ancestor amplitudes) counterterms, so that it appears like that the RG equation does not mix up the WPC hierarchy\footnote{At the level of amplitudes with the explicit dependence on the pion legs things are in general much more complicated and one has to resort to the LFE in order to fix them in a way compatible with the symmetry of the theory.}. 
However this is not sufficient to prove compatibility with the WPC, as 
the redefinition of the renormalized parameters $\rho_i$ still spans
in principle at a given order $n$ {\it all} (infinite) integrated local monomials corresponding to divergent ancestor amplitudes and compatible with Lorentz and SU(2)$_\s{R}$ global symmetry (under which $\tilde J_{a\mu}$ is in the adjoint and $K_0$ is a singlet) on the space of ancestor variables. What we need to prove is that $i$ spans only those monomials that are required by the WPC at order~$n$ and nothing else. 

\subsection{One-loop analysis}

To see where the problem resides and what need to be proven, let us consider the one-loop case $n=1$.
As already said, in this case there are five divergent ancestor amplitudes: $\G^{(1)}_{\widetilde J\widetilde J}$, $\G^{(1)}_{\widetilde J\widetilde J\widetilde J}$, $\G^{(1)}_{\widetilde J\widetilde J\widetilde J\widetilde J}$, $\G^{(1)}_{K_0 K_0}$, and $\G^{(1)}_{K_0\widetilde J\widetilde J}$. The one-loop topologies possibly contributing to these amplitudes are shown in~\fig{fig:JJ} through~\fig{fig:KJJ}.

The WPC-compatible tree-level couplings used to construct the ancestor amplitudes are the ones coming from the action~\noeq{tree-lev}. On the other hand, the effective action $\widehat\G^{(1)}$ will contain also the eight monomials~\noeq{theMs} with a coefficient\footnote{We reserve the notation $\la{i}$ (respectively $\lab{i}$) for the renormalized (respectively bare) parameters that are bound to be zero at tree-level due to the WPC. This is to be contrasted with $\rho_i$ (respectively $\rhob{i}$) which denote the renormalized (respectively bare) coefficients of {\em all} monomials (that is, including those that are non-zero according to the WPC).} $\lambda_i^{(1)}$, $i \neq 0$ (as already noticed,
the coefficient $\rho_0^{(1)}$ is zero at one loop level; in addition, notice that ${\cal M}_0$  is allowed by the WPC, so its coefficient is not of the $\lambda$-type), fixed by the divergent part of the corresponding ancestor amplitude. When trading the $K_0$ and $\widetilde J^a_\mu$ variables for the bleached ones, the monomials ${\cal M}_i$ will become the invariants ${\cal I}_i$ of~\1eq{theIs}; expanding then the bleached variables in terms of the pion fields as in~\2eqs{barK0exp}{jexp}, will generate new vertices 
with pions and external sources $\tilde J_\mu$ and $K_0$.
 Contrary to the tree-level vertices however, these so-called $\lambda$-vertices (see also the definition given in Sect.~\ref{genth}) violate the WPC and are proportional to a parameter $\lambda_i^{(1)}$, which was zero at tree-level (from which the name).

A Feynman graph constructed from this type of vertices might in principle contribute to the RG equation~\noeq{RGE} due to the derivative term in 
$\rho_i$, evaluated on the $\lambda$-type coefficients.
To understand how this can possibly happen, recall that for writing down the RG equation one writes down all possible tree-level couplings compatible with the LFE which are of two types: the coefficient $m_d$, which is also compatible with the WPC, and all the $\lambda$-vertices.
Notice that $g$ instead can be eliminated by redefining the pion fields $\phi_a \rightarrow \frac{1}{g} \phi_a$. The WPC condition selects the solution in which all the $\lambda_{i}$ are zero: it is on this solution that one evaluates the amplitude after taking the derivative wrt $\rho_{i}$ in~\1eq{RGE}. Consequently the insertion of more than one $\lambda$-vertex in an ancestor amplitude cannot contribute to the RG flow, since the WPC sets to zero all the $\lambda$ (the derivative obviously disposes of one such coefficient only). 

Thus, the compatibility of the RG flow with the WPC boils down to the proof that $\lambda$-vertices cannot contribute to the RG-flow of an ancestor amplitude.
As a warm up exercise in what follows we see how things work out at the one-loop order.

\subsubsection{$\widetilde J^a_\mu$ sector}

\begin{figure}[!t]
\centerline{\includegraphics[scale=0.7]{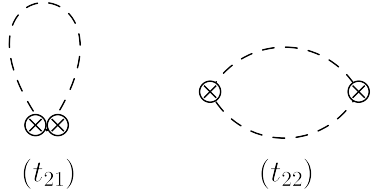}}
\caption{\label{fig:JJ}One-loop topologies contributing to the $\G^{(1)}_{\widetilde J\widetilde J}$ ancestor amplitudes. Crossed vertices indicate $\widetilde J$ external legs.}
\end{figure}

In the $\widetilde J^a_\mu$ sector the invariants that can contribute are ${\cal I}_1$ through ${\cal I}_5$ of~\1eq{theIs}. Let us then start by considering the two-point function $\Gamma^{(1)}_{\widetilde J\widetilde J}$, which has $D(\Gamma^{(1)}_{\widetilde J\widetilde J})=2$; the topologies possibly contributing to this amplitude are shown in ~\fig{fig:JJ}.

Now, all seagull type diagrams [indicated as $(t_{21})$ in \fig{fig:JJ}] vanish in dimensional regularization, as the pion has a massless propagator $\sim 1/k^2$. 
Consider next the topologies $(t_{22})$. According to our previous discussion, one of the vertices appearing there must come from the tree-level Feynman rules and it is given by
\begin{equation}
\frac{m_d^2}{4}\, \Tr (F_\mu - \widetilde J_\mu)^2 \sim -\frac{1}{2}g \epsilon_{abc} \widetilde{J}^a_\mu\partial^\mu \phi^b\phi^c,
\end{equation}
while the remaining vertex is a $\lambda$-vertex. Thus, we expand the first five invariants  in~\1eq{theIs} in powers of the pion fields keeping 
only terms of the form $\widetilde J\phi\phi$; one finds
\begin{align}
\mathcal{I}_1 &\sim -\frac{4}{m^2_d}g\epsilon_{abc}\left(2 \widetilde{J}^a_\mu \square \phi^b\partial^\mu \phi^c +
\partial^\mu \widetilde{J}^a_\mu\square \phi^b\phi^c\right),\nonumber\\
\mathcal{I}_2 &\sim -\frac{4}{m^2_d}g\epsilon_{abc} \left(
2 \widetilde{J}^a_\mu \partial^\mu \partial^\nu \phi^b \partial_\nu \phi^c
+\partial_\mu \widetilde{J}^a_\nu\partial^\mu \partial^\nu \phi_b\phi_c
+ \partial_\mu \widetilde{J}^a_\nu\partial^\mu \phi^b\partial^\nu \phi^c \right),\nonumber\\
\mathcal{I}_3 &\sim
-\frac{4}{m^2_d}\epsilon_{abc} \partial_\mu \widetilde{J}^a_\nu\partial^\mu \phi^b\partial^\nu \phi^c,
\nonumber\\
\mathcal{I}_4 &\sim 0;\qquad \mathcal{I}_5\sim 0.
\label{e.18}
\end{align}
Thus, all of the $\lambda$-vertices of this kind contain at least two derivatives acting on pion fields, being a rather remarkable fact that all potentially ``dangerous'' monomials (\ie monomials possessing only one or no derivatives acting on the pion fields) cancel out. As a consequence the UV degree of divergence of the topologies $(t_{22})$ is at least 3: Indeed, they all have one-loop, two bosonic propagators, one derivative from the tree-level vertex and at least two derivatives from the $\lambda$-vertex. Hence, since $D(t_{22})>D(\Gamma^{(1)}_{\widetilde J\widetilde J})$
all graphs of this kind cannot appear in the set of one-loop counterterms. 

\begin{figure}[!t]
\centerline{\includegraphics[scale=0.7]{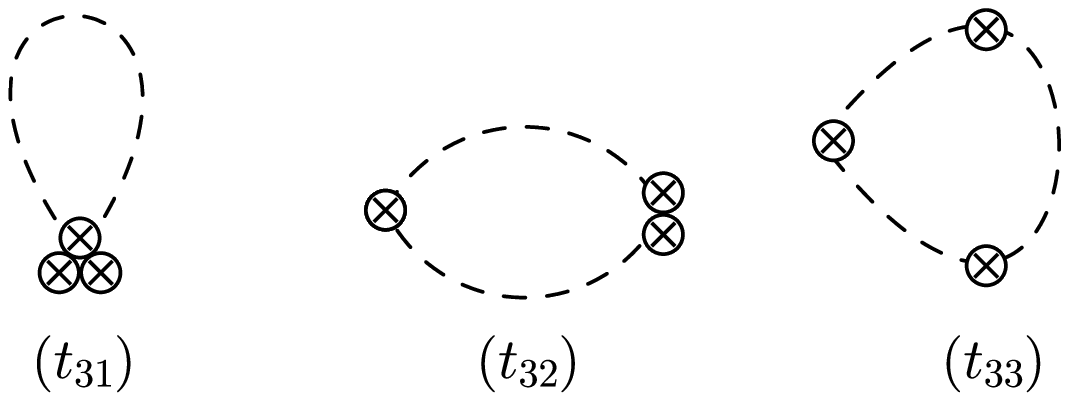}}
\caption{\label{fig:JJJ}One-loop topologies contributing to the $\G^{(1)}_{\widetilde{J}\widetilde{J}\widetilde{J}}$ ancestor amplitude.}
\end{figure}

We next consider the ancestor amplitude with three external ${\widetilde J}$ legs $\G^{(1)}_{\widetilde J\widetilde J\widetilde J}$, in which case $D(\Gamma^{(1)}_{\widetilde J\widetilde J\widetilde J})=1$; \fig{fig:JJJ} shows the possible topologies contributing to this amplitude.

As before the topology $(t_{31})$ vanishes; for graphs of the type $(t_{32})$ one has instead that the vertex with two external sources $\widetilde{J}$ and two pion fields has to be a $\lambda$-vertex since this kind of vertices is not present in the tree-level Feynman rules. Therefore, we now expand the five invariants~\noeq{theIs} in powers of $\phi$ keeping only terms of the form $\widetilde{J} \widetilde{J} \phi\phi$; we obtain
\begin{align}
\mathcal{I}_1 & \sim  \frac{4}{m^2_d}g^2\left(\widetilde{J}^a_\mu \widetilde{J}^a_\nu \partial^\mu \phi^b \partial^\nu \phi^b -
\widetilde{J}^a_\mu \widetilde{J}^b_\nu \partial^\mu \phi^b \partial^\nu \phi^a - \partial^\mu \widetilde{J}^a_\mu \widetilde{J}^b_\nu
\partial^\nu \phi^a \phi^b + \partial^\mu \widetilde{J}^a_\mu \widetilde{J}^b_\nu\,\partial^\nu \phi^b \phi^a\right),\nonumber\\
\mathcal{I}_2 &\sim
\frac{4}{m^2_d}g^2\left(\widetilde{J}^2 \partial_\mu \phi_a \partial^\mu \phi^a -
\widetilde{J}^a_\mu \widetilde{J}^{b\mu} \partial_\nu \phi_a \partial^\nu \phi^b -\partial_\mu \widetilde{J}^a_\nu \widetilde{J}^{b\nu}
\partial^\mu \phi^a \phi^b + \partial^\mu \widetilde{J}^a_\nu \widetilde{J}^{b\nu}\partial^\mu \phi^b \phi^a\right),
\nonumber\\
\mathcal{I}_3 &\sim \frac{2}{m^2_d}g\Big(3\widetilde{J}^a_\mu  \widetilde{J}^b_\nu\, \partial^\mu \phi^a \partial^\nu \phi^b
+ 2 \widetilde{J}^2 \partial_\mu \phi^a \partial^\mu \phi^a - 2\widetilde{J}^a_\mu  \widetilde{J}^a_\nu \partial^\mu \phi^b \partial^\nu \phi^b
-2\widetilde{J}^a_\mu  \widetilde{J}^{b\mu}\partial_\nu \phi_a \partial^\nu \phi_b
\nonumber\\
& - \widetilde{J}^a_\mu  \widetilde{J}^b_\nu \partial^\mu \phi_b \partial^\nu \phi_a + \partial_\mu  \widetilde{J}^a_\nu
 \widetilde{J}^{b\mu} \partial^\nu \phi_a \phi_b -\partial_\mu \widetilde{J}^a_\nu \widetilde{J}^{b\mu}\partial^\nu \phi_b \phi_a + 
\partial_\mu  \widetilde{J}^a_\nu \widetilde{J}^{b\nu}\partial^\mu \phi_b \phi_a
\nonumber\\
& -\partial_\mu  \widetilde{J}^a_\nu \widetilde{J}^{b\nu} \partial^\mu \phi_a \phi_b\Big),\nonumber\\   
\mathcal{I}_4 & \sim \frac{8}{m^2_d}\left(\widetilde{J}^2 \partial_\mu \phi_a  \partial^\mu \phi_a + 2 \widetilde{J}^a_\mu \widetilde{J}^b_\nu \partial^\mu \phi_a 
\partial^\nu \phi_b\right), \nonumber\\
\mathcal{I}_5& \sim \frac{8}{m^2_d} \left(\widetilde{J}^a_\mu \widetilde{J}^a_\nu  \partial^\mu \phi_b \partial^\nu \phi_b +
\widetilde{J}^a_\mu  \widetilde{J}^b_\nu \partial^\mu \phi_b \partial^\nu \phi_a +
\widetilde{J}^a_\mu  \widetilde{J}^{b\mu}  \partial_\nu \phi_a \partial^\nu \phi_b\right).
\label{e.19}
\end{align}
Notice that the above vertices contain at least one derivative acting on a pion field. Therefore, the UV degree of divergence of the $(t_{32})$ graphs  is at least 2 (the power counting is the same as that of graphs $(t_{22})$ in \fig{fig:JJ} apart for the fact that in this case the $\lambda$-vertex contains at least one derivative). Thus, $D(t_{32})>D(\Gamma^{(1)}_{\widetilde J\widetilde J\widetilde J})$ so that, as in the previous case, this kind of graphs cannot contribute to one-loop counterterms. 

Finally, the UV degree of divergence of $(t_{33})$ graphs must be at least 2, since they all have one loop, three bosonic propagators, one derivative from every tree-level vertex and at least two derivatives from the $\lambda$-vertex. So, by the same token, also these diagrams do not appear in the one-loop RG equation for $\Gamma^{(1)}_{\widetilde{J}\widetilde{J}\widetilde{J}}$.

\begin{figure}[!t]
\centerline{\includegraphics[width=0.8\textwidth]{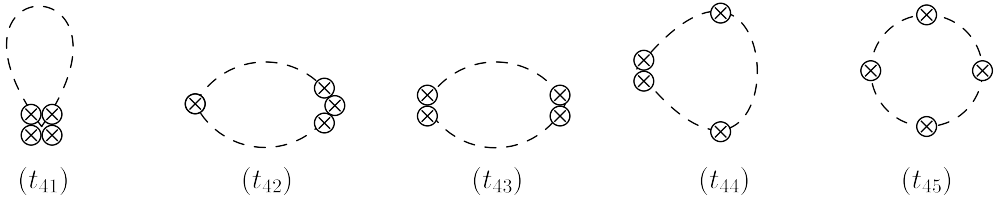}}
\caption{\label{fig:JJJJ}One-loop topologies contributing to the $\Gamma^{(1)}_{\widetilde{J}\widetilde{J}\widetilde{J}\widetilde{J}}$ ancestor amplitude.}
\end{figure}

The last ancestor amplitude is the one containing four external $\widetilde J$ sources, in which case $D(\Gamma^{(1)}_{\widetilde J\widetilde J\widetilde J\widetilde J})=0$. The topologies contributing to such amplitude are finally shown in~\fig{fig:JJJJ}.

Once again the massless seagull graphs $(t_{41})$ vanish. Next, in order to compute the UV degree of divergence of the topology $(t_{42})$, we need to expand 
the one-loop invariants in powers of $\phi$ keeping 
only terms of the form $\widetilde{J} \widetilde{J} \widetilde{J} \phi\phi$. One finds
\begin{align}
\mathcal{I}_1 &\sim 0; \qquad \mathcal{I}_2 \sim 0,\nonumber\\
\mathcal{I}_3 &\sim \frac{2}{m^2_d}g^2\epsilon_{abc}\Big(\widetilde{J}^d_\mu \widetilde{J}^d_\nu \widetilde{J}^{a\nu} \, \partial^\mu \phi^b\, \phi^c 
- \widetilde{J}^2 \widetilde{J}^a_\mu \, \partial^\mu \phi^b\, \phi^c\Big),
\nonumber\\
\mathcal{I}_4 &\sim -\frac{8}{m_d^2} g \epsilon_{abc}\widetilde{J}^2\, \widetilde{J}^a_\mu \partial^\mu \phi_b\, \phi_c, \nonumber\\
\mathcal{I}_5 &\sim -\frac{8}{m_d^2}g \epsilon_{bcd}\, \widetilde{J}^a_\mu \widetilde{J}^a_\nu \widetilde{J}^{b\nu}\partial^\mu \phi_c \phi_d.
\label{e.20}
\end{align}
Notice that also the vertices of this kind contain one derivative acting on a pion field. However $(t_{42})$ graphs 
have an UV degree of divergence which is at least one irrespectively of the number of derivatives in the $\lambda$-vertex. Thus, $D(t_{42})>0$ and therefore the one-loop invariants with four external sources $\widetilde{J}$, cannot receive contributions from these Feynman diagrams.

Also $(t_{43})$ graphs do not appear in the one-loop RG equation for $\Gamma^{(1)}_{\widetilde{J}\widetilde{J}\widetilde{J}\widetilde{J}}$ because in the tree-level 
Feynman rules there are no vertices of the form $\widetilde{J} \widetilde{J} \phi \phi$ and so in this topology both vertices must necessarily be of the WPC violating type. Finally, using the previous results, it is straightforward to prove that the UV degree of divergence of the graphs of type $(t_{44})$
(three bosonic propagators, two tree-level vertices with one derivative and a $\lambda$-vertex with at least one derivative) and $(t_{45})$  
(four bosonic propagators, three tree-level vertices with one derivative and a $\lambda$-vertex with at least two derivatives) is at least one, 
so that also in this case $D(t_{44})\,, D(t_{45})>0$.       

This completes the analysis of the $\widetilde{J}$-sector at the one-loop level and shows that for the two-, three- and four-point functions of the external source $\widetilde{J}$ a change in the scale $\mu$ is compensated by a change of the finite parts of genuinely one-loop invariants. 
Let us conclude, by observing that even though we have taken into account only $\lambda$-vertices stemming from the one-loop invariants~\noeq{theIs}, the argument is valid also, {\it a fortiori}, for monomials that appear as counterterms at higher loops, for the latter will contain either more derivatives or more bleached variables $j$ (or both).

\subsubsection{$K_0$-sector}

In the $K_0$ sector one needs to consider the invariants ${\cal I}_6$ and ${\cal I}_7$ of~\1eq{theIs}.
However, recall that the tree-level dependence from $K_0$ is completely fixed by the coupling to the non-linear constraint $\phi_0$. Indeed the non-linear symmetry is realized through the transformations~\noeq{nltrans} and the only composite operator that enters in them is $\phi_0$. This dictates the coupling in $S_\s{\rm ext}$ of~\1eq{tree-lev}, and it makes no sense to insert at tree-level additional invariants that depend on $K_0$. Thus there are no $\lambda$-vertices originating from invariants involving $\overline{K}_0$.

This means in turn that there are no $\lambda$-vertices contributing to the two point function of the scalar source $K_0$; hence, a change in the scale $\mu$ in the ancestor amplitude $\Gamma^{(1)}_{K_0 K_0}$ can be compensated by a change of the 
finite part of the one-loop counterterm $\mathcal{I}_4$.

\begin{figure}[!t]
\begin{center}
\includegraphics[scale=0.7]{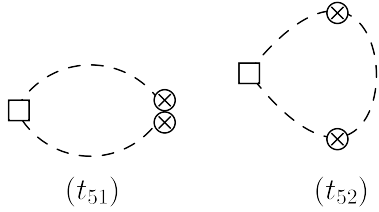}
\end{center}
\caption{\label{fig:KJJ}One-loop topologies contributing to the $\G^{(1)}_{K_0 \widetilde{J}\widetilde{J}}$ ancestor amplitude. Square vertices indicate $K_0$ external legs.}
\end{figure}

Finally, we show in~\fig{fig:KJJ} the topologies that contribute to the ancestor amplitude with one external scalar source $K_0$ and two $\tilde{J}$ legs.

As discussed above, the $\lambda$-vertex in both topologies can never come from the $K_0$ source. Therefore for $(t_{51})$ graphs the $\lambda$-vertex contains at least one derivative; hence the UV degree of divergence of these graphs is at least 1. 
This is also the UV degree of divergence for the topologies $(t_{52})$. Therefore, we conclude that the one-loop invariant with one scalar source $K_0$ and two $\widetilde{J}$, which does not contain derivatives, cannot receive contributions from these 
graphs.

Also in this sector the results obtained are valid for $\lambda$-vertices originating from higher order counterterms.
This completes the analysis of the $K_0$-sector and it allows us to conclude that at the one-loop level a change in the scale $\mu$ only requires a change in 
the finite parts of the WPC one-loop invariants to be compensated. 

\section{\label{genth}A General theorem}

We are now ready to tackle a general proof of the fact that the WPC is preserved by the RG-flow. The strategy followed for proving this will be different from the one adopted for illustrating the one-loop case, which obviously cannot be adapted to an all-order analysis. 

Before dwelling on the detailed proof let us recall the precise definition of a $\lambda$-vertex and state in a precise form the theorem we would like to prove.

\medskip

\noindent{\bf Definition}. A $\lambda$-vertex is an interaction vertex generated upon the expansion of a symmetric (\ie fulfilling the LFE) local functional forbidden by the WPC in powers of the pion fields $\phi$. It must contain at least one external $\widetilde J$ leg\footnote{Again we remind that the dependence of the tree-level effective action on $K_0$ is fixed by the nonlinear SU(2) symmetry and hence no $\lambda$-vertices originating from invariants involving $\overline{K}_0$ need to be considered.}.

\medskip 

\noindent One has then the following

\medskip
\noindent{\bf Theorem}. There are no $\lambda$-vertex contributions to the RG-flow of a $n^{\rm{th}}$ loop ancestor amplitude. 
\medskip

\noindent The proof is divided into several steps that we detail in the following five subsections.

\subsection{Loop expansion of the RG equation}

Let us consider a $\lambda_i$ parameter that, according to the WPC,
is bound to be zero up to the order $n$:
\begin{equation}
\lambda_i = \lambda_i^B + O(\hbar^n),
\label{loop.1}
\end{equation}
and suppose that there exists an order $m<n$ such that a contribution to the RG equation arises at that order from the $\lambda_i$-parameter. 

Such a  contribution to the second term in the lhs of the RG equation (\ref{RGE}) for the ancestor amplitudes is given by
\begin{equation}
\left . \mu \frac{\partial \lambda_i^{(m)}}{\partial \mu}
\frac{\partial \G_0^{(0)}}{\partial \lambda_i^B} \right |_{\lambda=0} = 
 \left . \mu \frac{\partial \lambda_i^{(m)}}{\partial \mu} \right |_{\lambda=0}
 {\cal M}_i.
\label{loop.2}
\end{equation}

Notice that one obtains a local contribution, as it should be 
since $m$ is the lowest order where $\lambda_i$ is assumed
to contribute.
If such an integer $m<n$ existed, one would clearly
mix up the WPC counting: a change in the scale $\mu$ would be reflected
in lower order contributions, associated with counterterms that 
cannot appear at that order according to the WPC.

Let us now prove that this is indeed not the case.

\subsection{Topologies}

\begin{figure}[!t]
\centerline{\includegraphics[scale=0.7]{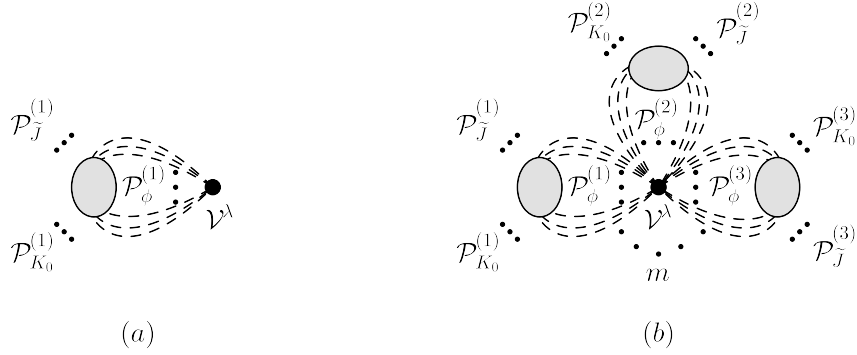}}
\caption{\label{fig:daisy}Relevant topologies for the RG-flow theorem. Graph $(a)$ shows a single petal diagram, which, once iterated, gives rise to the multi-petal daisy diagram $(b)$. To avoid notational cluttering we do not indicate explicitly the $\widetilde J_\mu^a$ and $K_0$ vertices either in the ${\cal P}$-amplitude or in $\vlam$.}
\end{figure}

The amplitudes that we need to consider have no external pion legs and display an insertion of a single $\lambda$-vertex $\vlam$, all other vertices being generated by the WPC tree-level action~\noeq{tree-lev}. We will denote with $\vlamK$ (respectively, $\vlamJ$) the number of $K_0$ (respectively, ${\widetilde J}^a_\mu$) legs attached to the vertex $\vlam$.  Finally $r$ will denote the number of pion legs attached to $\vlam$.

The relevant $n^{\rm{th}}$-loop topologies can be classified according to the number of petals ${\cal P}$ composing a daisy diagram centered on $\vlam$, see~\fig{fig:daisy}. These ${\cal P}$-amplitudes are descendant amplitudes obtained from the $n^{(i)}$-order ancestors after writing them in terms of the bleached variables plus (whenever $n^{(i)}\geq2$) the contribution of the functional ${\cal G}$ of~\1eq{fullLFEsol}.
They correspond to all possible partitions of the integers $1,\dots,r$ in disjoint sets each of which has at least 2 elements; in particular, if $n^{(i)}$ is the loop order of the $i$-th petal amplitude, then, since $r^{(i)}$ propagators give rise to $r^{(i)}-1$ loops, one has obviously
\begin{equation}
n=\sum_i n^{(i)}+\sum_i [r^{(i)}-1],
\end{equation}
or $n=n^{(1)}+r-1$ for just one petal. 

\subsection{${\cal P}$-amplitudes degree of divergence}

Let us indicate with $\Piphi{i}$ (respectively,  $\PiK{i}$, $\PiJ{i}$)  the number of $\phi$ (respectively, $K_0$, ${\widetilde J}^a_\mu$) legs attached to the 1PI amplitude building up the $i$-th petal $\Pi{i}$. Then, one has the following

\medskip

\noindent {\bf Lemma}. The degree of divergence of a ${\cal P}^{(i)}$-amplitude satisfies the WPC bound~\noeq{WPC}, \ie
\begin{equation}
D({\cal P}^{(i)})=(d-2)n^{(i)}+2-\PiJ{i}-2\PiK{i}.
\label{Pbound}
\end{equation}

\medskip

To prove this, let us analyze the two possible contributions to a ${\cal P}^{(i)}$-amplitude, that is the one coming from the bleached variables substitution and the one from the ${\cal G}$ term of the general solution of the LFE.

Let 
\begin{equation}
{\cal A}^{(i)}_{\underbrace{\widetilde J\cdots\widetilde J}_{N_{\widetilde J}}\underbrace{K_0\cdots K_0}_{N_{K_0}}\underbrace{\widetilde J\cdots\widetilde J}_{\PiJ{i}}\underbrace{K_0\cdots K_0}_{\PiK{i}}},
\label{ancP}
\end{equation}
be an ancestor amplitude that, upon the substitution of the $N_{K_0}$ and $N_{\widetilde J}$ legs contributes to the ${\cal P}^{(i)}$-amplitude under scrutiny
\begin{equation}
{\cal P}^{(i)}_{\underbrace{\phi\cdots\phi}_{r^{(i)}}\underbrace{\widetilde J\cdots\widetilde J}_{\PiJ{i}}\underbrace{K_0\cdots K_0}_{\PiK{i}}}.
\label{descP}
\end{equation}
Then
\begin{equation}
D({\cal A}^{(i)})=(d-2)n^{(i)}+2-(N_{\widetilde J}+\PiJ{i})-2(N_{K_0}+\PiK{i}).
\end{equation}

Now, according to their definition~\noeq{bleached} and the corresponding expansions~\noeq{barK0exp} and~\noeq{jexp}, one observes that each substitution of a $\widetilde J$-leg can give at most one derivative acting on the pion fields,
while in the case of a $K_0$-leg 
 one gets at most two derivatives. Then one has
\begin{align}
D({\cal P}^{(i)})&=D({\cal A}^{(i)})+\hspace{-1.05cm}\underbrace{1\times N_{\widetilde J}}_{{\rm one}\ \partial\ {\rm for\ each\ replaced\ } \widetilde J}\hspace{-1.05cm}+\hspace{-1.05cm}\overbrace{2\times N_{K_0}}^{{\rm two}\ \partial{\rm 's\ for\ each\ replaced\ } K_0}\nonumber \\
&=(d-2)n^{(i)}+2-\PiJ{i}-2\PiK{i}.
\end{align}

However this does not exhaust all the possible ancestor amplitudes eventually contributing to the given ${\cal P}$-amplitude. In fact, observe that single derivatives of the pion fields coming from the replacement $K_0\to\overline{K}_0/m_d$ get accompanied by an extra $\widetilde J^a_\mu$; the same is true for zero derivative pion fields terms coming from the replacement $\widetilde J^a_\mu\to -j^a_\mu$. Thus one has the following two cases:

\begin{itemize}

\item[\n{i}] If we start from an ancestor amplitude and in the replacement of a $\widetilde J$ we get no derivative and a $\widetilde J$, the number of external $\widetilde J$ of the descendant amplitude stays invariant;

\item[\n{ii}] If we start from an ancestor amplitude and in the replacement of a $K_0$-leg we get a derivative and a $\widetilde J$, then the additional contribution to the degree of divergence is through the term
\begin{equation}
\underbrace{-2}_{{\rm from\ the}\ -2N_{K_0} {\rm \ term\ of\ } D({\cal A}^{(i)})}\hspace{-3cm}\overbrace{+1}^{{\rm from\ the\ derivative}}\hspace{-1cm}=-1
\end{equation}
that is we get exactly the contribution of the additional $\widetilde J$-leg generated. 

\end{itemize}

Thus we conclude that for the part of the petal amplitude that is generated through the bleached variables the bound~\noeq{Pbound} holds.

Next consider the contribution coming from the bilinear term of the LFE. We know that the bound is satisfied when $n=1$, so that we can proceed by induction. We then consider the differential operator
\begin{equation}
{\cal O}=\frac{\delta^{\ell+s+t}}{\delta \widetilde J^{a_1}_{\mu_1}(x_1)\cdots \delta \widetilde J^{a_\ell}_{\mu_\ell}(x_\ell)\delta K_0(y_1)\cdots\delta K_0(y_s)\delta\phi^{b_1}(z_1)\cdots\delta\phi^{b_t}(z_t)}
\end{equation}
and we apply it to the bilinear equation~\noeq{bilieq}. This operation will give rise to a term contributing to an amplitude involving $\ell$ $\widetilde J$-legs, $s$ $K_0$-legs and $t+1$ $\phi$-legs. As there are many possible ways of acting on the rhs of~\1eq{bilieq}, we then denote by $n_{\widetilde J}^\s{(\rm I)}$ and $n_{\widetilde J}^\s{(\rm II)}$ the number of $\widetilde J$ derivatives acting on $\frac{\delta\G^{(j)}}{\delta K_0(x)}$ and $\frac{\delta\G^{(n-j)}}{\delta \phi^a(x)}$ respectively, and similarly for $n_{K_0}^\s{(\rm I)}$ and $n_{K_0}^\s{(\rm II)}$. 

Using the induction hypothesis, the UV degree of divergence of the two amplitudes obtained in this way are 
\begin{align}
D({\rm I})&=(d-2)j+2-n_{\widetilde J}^\s{(\rm I)}-2n_{K_0}^\s{(\rm I)}-2,\nonumber \\
D({\rm II})&=(d-2)(n-j)+2-n_{\widetilde J}^\s{(\rm II)}-2n_{K_0}^\s{(\rm II)}.
\end{align}
Summing everything up one gets again the result
\begin{equation}
D({\cal O})=(d-2)n+2-\ell-2s,
\end{equation}
that is the bound holds at order $n$.

This concludes the proof of the lemma.

\subsection{One-petal amplitudes}

Consider now the $n^{\rm th}$-order amplitude $\G$ corresponding to a $\lambda$-vertex, with $r$ $\phi$-legs, $\vlamJ$ $\widetilde J$-legs and $\vlamK$ $K_0$-legs. Assume also that there is a single petal ${\cal P}^{(1)}$ (see diagram $(a)$ in~\fig{fig:daisy}), so that $r\equiv\Piphi{1}$ and $n=n^{(1)}+\Piphi{1}-1$.

The UV degree of divergence of this amplitude is
\begin{equation}
D(\G)=d(\Piphi{1}-1)-2\Piphi{1}+D({\cal P}^{(1)})+\delta,
\end{equation}
where $\delta\ge0$ is the number of derivatives present in the $\lambda$-vertex.
Since $n^{(1)}=n-r+1$, by means of the lemma previously proved we can  write
\begin{equation}
D(\Gamma)=(d-2)n+\delta+\PiJ{1}-2\PiK{1}.
\end{equation}

A bound on $\delta$ can be then obtained by observing that the degree above cannot be greater than the one of a $n^{\rm th}$-order  ancestor 
amplitude with $\vlamJ+\PiJ{1}$ $\widetilde J$-legs and $\vlamK+\PiK{1}$ $K_0$-legs, that is
\begin{equation}
D(\G)\le(d-2)n+2-(\PiJ{1}+\vlamJ)-2(\PiK{1}+\vlamK),
\end{equation}
thus we get the inequality
\begin{equation}
0\le\delta\le2-\vlamJ-2\vlamK.
\end{equation}
One has then the following cases
\begin{itemize}

\item[\n{i}] If $\vlamK=1$, then $\vlamJ=\delta=0$ (\ie the $\lambda$-vertex has no derivatives). In this case the $\lambda$-vertex is of the type $K_0\phi_0$, already present in $\G^{(0)}$ and allowed by the WPC.

\item[\n{ii}] If $\vlamK=0$, then either $\vlamJ=1$, in which case $\delta=1$ so that these are the couplings of the type $\widetilde J F$ coming from $S$, or $\vlamJ=2$, in which case $\delta=0$ so that these are the couplings of the type $\widetilde J^2$ coming again from $S$.

\end{itemize}

Thus we find that the only possible $\lambda$-vertices are the ones  allowed by the WPC, which proves the theorem at the level of a single petal amplitude.

\subsection{Daisy amplitudes}

Let us now consider a full daisy graph composed by $m$ ${\cal P}$-amplitudes. For such an amplitude one has then
\begin{equation}
D(\G)=\sum_{i=1}^m[d(\Piphi{i}-1)-2\Piphi{i}+D({\cal P}^{(i)})]+\delta',
\label{bmulti}
\end{equation}
where $m$ is the number of petals attached to the $\lambda$-vertex $\vlam$  and $\delta'$ is the number of derivatives of the $\lambda$-vertex.
The lemma tells us that for each one of the petal amplitudes ${\cal P}^{(i)}$ the degree of divergence is
\begin{align}
D({\cal P}^{(i)})=(d-2)n^{(i)}+2-\PiJ{i}-2\PiK{i}.
\end{align}
In addition, the loop order of the ${\cal P}^{(m)}$ amplitude is given by
\begin{equation}
n^{(m)}=n+m-\Piphi{m}-\sum_{i=1}^{m-1}[n^{(i)}+\Piphi{i}].
\end{equation}
Then, after simple algebra,~\1eq{bmulti} yields
\begin{equation}
D(\Gamma)=(d-2)n+\delta'-\sum_{i=1}^m[\PiJ{i}+2\PiK{i}].
\end{equation}
At this point one obtains the same bound as before on $\delta'$, and therefore the same conclusions hold.
\medskip

\noindent This completes the proof of the RG-flow theorem.

\section{\label{stability}Stability in WPC Renormalizable Theories}

The existence of a RG equation allows one to extend the notion of stability of the classical theory to the non-linearly realized models based on the WPC.

In fact, the WPC prescribes uniquely which coefficients $\rho_i$ are non-zero at tree-level (and therefore it defines the set of $\lambda$-invariants). In addition, the RG flow theorem just proven implies that the finite parts of the counterterms, needed to reabsorb a change in the $\mu$-dependence, appear exactly at the order where the pole part of the corresponding ancestor amplitudes becomes non-zero according to the WPC.

Suppose now that one adds a free finite ($\mu$-independent) part $a^{(n)}_i$ at order $n$ in the loop expansion, in a way to preserve the symmetries of the theory (and not violating the WPC). As the loop order $n$ of the finite coefficient $a^{(n)}_i$ does not correspond to the topological loop order\footnote{For example, a local counterterm added to remove a one-loop divergence corresponds topologically to a tree-level graph.}, one can proceed as in the power-counting renormalizable theories, that is one rescales $a^{(n)}_i \rightarrow \frac{1}{\hbar^n} a^{(n)}_i$ to obtain a finite physically equivalent theory (as $\hbar = 1$).
  
In the power-counting renormalizable case, the rescaling will give back a term already present at the tree-level: this is the well-known stability of the tree-level action against radiative corrections. For WPC renormalizable theories however, under the rescaling the addition of a finite free coefficient $a^{(n)}_i$ at the order $n$ in the loop expansion is equivalent to the 
addition of a non-zero $\lambda$-vertex at tree-level in the rescaled theory. For such vertices a change in the scale $\mu$ cannot be anymore accommodated by a change of a finite number of counterterms order by order in the loop expansion, and one would inevitably end mixing up the WPC criterion on the loop order of UV divergences in the rescaled theory.

Thus one can extend the notion of stability of the classical theory: If one demands that the rescaled   theory satisfies the WPC, then there is no freedom left to add any finite $\mu$-independent terms and the theory is (weakly) stable under radiative corrections. 

Incidentally, notice that there is yet another reason why one cannot add the symmetric finite renormalization terms $a_i^{(n)}$ just discussed. The WPC uniquely identifies the graphs in the expansion based on the topological loop number, thus selecting a particular Hopf algebra, as the latter is constructed as a dual of the enveloping algebra of the Lie algebra of the Feynman graphs associated to the theory under scrutiny~\cite{Connes:1999yr,Connes:2000fe}. On the other hand, it also guarantees that there exists a suitable exponential map on this Hopf algebra~\cite{EbrahimiFard:2010yy} which allows the removal of all the divergences.
The addition of any $a_i^{(n)}$ is equivalent to a change in the Hopf algebra of the model, as it would modify the set of 1-PI Feynman diagrams on which the Hopf algebra is constructed.  This change destroys the compatibility between the WPC and the RG equation; therefore, the addition of such terms is not allowed. 

The existence of a connection between the WPC preserving RG flow and the selected Hopf algebra clearly deserves further investigations.

\section{\label{bsm}Beyond the Standard Model: the WPC as a model-building principle}

If one promotes the classical source $\widetilde J^a_\mu$ to the status of a dynamical field, the \nlsm action gives rise to the St\"uckelberg mass term. By formulating a non-linearly realized SU(2) Yang-Mills theory in the LFE framework \cite{Bettinelli:2007tq,Bettinelli:2007cy,Bettinelli:2007eu} (with the pseudo-Goldstone fields taking over to the role of the pion fields) and imposing the WPC, one arrives to a somewhat surprising result. 

Specifically, notwithstanding the fact that if the local gauge symmetry is realized non-linearly the Yang-Mills action is not singled out on the basis of gauge invariance\footnote{In particular, all possible monomials constructed from the bleached variable $j^a_\mu$ and its ordinary derivatives, are gauge invariant, and therefore can in principle appear as interaction vertices in the classical action.}, it turns out that if the WPC condition is satisfied, then the only admissible solution of the tree-level LFE is the Yang-Mills action plus the St\"uckelberg mass term:
\begin{align}
S_\s{\rm nlYM}&=S_\s{\rm YM}+\frac{M^2}2\int\!\diff{d}x\,(A_\mu^a-F_\mu^a)^2;&
S_\s{\rm YM}&=-\frac{1}{4}\int\!\diff{d}x\,G^a_{\mu\nu}G_a^{\mu\nu},
\end{align}
where $G^a_{\mu\nu}$ is the field strength of the gauge field $A_\mu^a$ and $F^a_\mu$ the flat connection.

An important consequence of this fact is that one can formulate a non-linearly realized theory based on the gauge group SU(2)$\times$U(1) 
\cite{Bettinelli:2008ey, Bettinelli:2008qn,Bettinelli:2009wu}. One combines the SU(2) gauge fields $A_\mu = A_{\mu}^a \frac{\tau_a}{2}$ and
the $\rm U(1)_\s{Y}$ gauge field $B_\mu$ into the bleached variable
\begin{align}
w_\mu &= w_{a\mu} \frac{\tau_a}{2}= \Omega^\dagger g A_\mu \Omega + g' \frac{\tau_3}{2} B_\mu + i \Omega^\dagger \partial_\mu \Omega.
\label{a.g.1}
\end{align} 
$g$ and $g'$ are the SU(2) and U(1) coupling constants respectively.
The bleached counterparts of the $A$, $Z$ and $W^\pm$ fields are given by
\begin{align}
A_\mu &= - \sw A_{3\mu} + \cw B_\mu;&
Z_\mu &= \left . \frac{1}{\sqrt{g^2 + g'^2}} w_{3\mu} \right |_{\phi_a = 0} =\cw A_{3\mu} + \sw B_\mu,\nonumber \\
w^\pm_\mu &= \frac{1}{\sqrt{2}} ( w_{1\mu} \mp i w_{2\mu} ),
\end{align}
where $\sw$ ($\cw$) is the sine (cosine) of the Weinberg angle the tangent of which is given by the ratio $g'/g$. Bleached fermions are obtained by left multiplication with $\Omega^\dagger$ of the original doublet.
For a generic SU(2) doublet $L$, its bleached counterpart is 
\begin{align}
\tilde L = \Omega^\dagger L \, .
\label{bleached.fermion}
\end{align}
The action of this non-linear version of the Eletroweak Theory
is highly constrained if one requires the WPC to be satisfied: in this case the self-couplings of gauge bosons as well as the couplings between gauge bosons and fermions are the same as the conventional SM ones. However, the following combination of two independent mass invariants arise, and the Weinberg relation is broken
\begin{align}
&M^2_\s{W}w^+w^-+\frac{M^2_\s{Z}}2w^2_{3}; & 
M_\s{Z}^2 &= (1 + \kappa) \frac{M_\s{W}^2}{\cw^2}.
\label{gauge.boson.masses}
\end{align}
This is a peculiar feature of nonlinearly realized electroweak theories~\cite{Quadri:2010uk}.

The inclusion of physical scalar resonances in the non-linearly realized Electroweak Theory, while respecting the WPC, yields a definite beyond the Standard Model (bSM) scenario. Indeed it turns out that it is impossible to add a scalar singlet without breaking the WPC condition~\cite{Binosi:2012cz}. The minimal solution requires a SU(2) scalar physical doublet, leading to a CP-even physical field (to be identified with the recently discovered scalar resonance at 125.6 GeV) and three additional heavier physical states, one neutral CP-odd and and two charged ones. Notice that this is a rather peculiar signature, since in Two Higgs-Doublet Models and the Minimal Supersymmetric Standard Model the number of physical scalar resonances is five (see, \eg~\cite{Gunion:1989we}).

While some preliminary phenomenology issues of this model have been addressed in~\cite{Binosi:2012cz}, a full analysis and comparison with the experimental data can be carried out in a satisfactory and systematic way only in the presence of an RG-flow, as the possibility of running the scale $\mu$ in a mathematically consistent way would allow one to obtain physical predictions of the same observables applicable in different energy regimes.

The extension of the analysis carried out for the \nlsm to the non-linearly realized electroweak theory with scalar resonances (NLSM for short) requires some care.

The mass generation mechanism for gauge bosons cannot be entirely of the St\"uckelberg type, for in this case the decays of the Higgs scalar $h \rightarrow WW^*$ and $h \rightarrow ZZ^*$ would be loop-induced and thus phenomenologically unacceptably small. Therefore  the problem arises to assess whether a fraction of the mass is generated by the St\"uckelberg mechanism, in addition to the contribution associated with the linearly realized spontaneous symmetry breaking {\em \`a la} Higgs.

Moreover, since current LHC data are in very good agreement with the Standard Model~\cite{Giardino:2013bma,Ellis:2013lra}, one can assume in a first approximation custodial symmetry and set $\kappa =0$ in ~\1eq{gauge.boson.masses}. The $W$ and $Z$ masses are therefore
\begin{equation}
M_\s{W} = \frac{gv}{2} \sqrt{1 + \frac{A}{v^2}}; \qquad
M_\s{W} = \frac{Gv}{2} \sqrt{1 + \frac{A}{v^2}},
\label{W.and.Z.masses}
\end{equation}
where $G = \sqrt{g^2 + {g'}^2}$ and $A$ is a parameter of mass
dimension squared describing the fraction of the mass generated
by the St\"uckelberg mechanism.

The Lagrange multiplier formulation of the non-linear constraint is particularly suited for studying the small $A$ limit, which is the phenomenologically relevant regime since  bSM effects are known to be small~\cite{Giardino:2013bma,Ellis:2013lra}. This is because in this formulation it is easy to derive the dominant contribution in the small $A$ expansion of the 1-PI amplitudes (without the need of resummations)~\cite{Bettinelli:2013hia}.

The Lagrange multiplier (BRST-invariant)
implementation of the non-linear constraint
\begin{align}
S_\s{\rm embed} = \int\! {\rm d}^4x \, s(\bar c B) = \int\! {\rm d}^4x \, \left\{
B\left[( \sigma+f)^2 + \phi_a^2 - f^2\right] - \bar c c \right\}
\end{align}
is realized by introducing a pair of BRST variables
$B,c$ such that $sB = c$, $sc = 0$. $B$ is the Lagrange multiplier field, $c$ is the associated ghost (that is free). On the other hand, the BRST variation of the antighost $\bar c$ yields the (invariant) non-linear constraint:
\begin{align}
s \bar c = (\sigma + f)^2 + \phi_a^2 - f^2,
\end{align}
where $f$ is the mass parameter expressing the v.e.v.\footnote{In the NLSM model, $v$ is reserved for the v.e.v. induced by the {\em linear} spontaneous symmetry breaking mechanism.} of the trace component $\phi_0$ of the SU(2) matrix $\Omega$. 

There is no WPC in the sector spanned by $\bar c$ and its BRST variation $s \bar c$~\cite{Bettinelli:2013hia}. However, since they form a BRST doublet\footnote{A BRST doublet is a set of variables $u$ and $v$ transforming under the BRST operator $s$ according to $su=v$, $sv=0$.}~\cite{Quadri:2002nh}, it turns out that they can only modify the BRST-exact sector of the theory, which, like the gauge-fixing, is not physical.

On the other hand, in the gauge-invariant (BRST-closed) sector of the theory, the WPC holds. In  this sector the proof of the compatibility between the WPC and the RG flow given in Sect.~\ref{genth} can be extended to the NLSM. Indeed all quantized fields and external sources of the model transform under the nonlinearly realized gauge group either as a connection, or in the fundamental representation (like the fermions) or else in the adjoint representation (like, for instance, the ghost fields). 
It turns out that the bleaching procedure for a variable that transforms in the fundamental or in the adjoint representation of the gauge group does not involve derivatives. To be sure, one has the expressions~$\widetilde f = \Omega^\dagger f$ (respectively, $\widetilde X = \Omega^\dagger X \Omega$) for the bleached variable associated to a field that transforms in the fundamental (respectively, adjoint) representation of the gauge group. Therefore the substitution of these fields with their bleached counterparts does not modify the degree of divergence of the amplitude considered.

Moreover one should also note that in a massive theory the WPC provides only an upper bound on the degree of divergence of an ancestor amplitudes, with the 
bound being saturated only in the massless case. 
Yet the upper bound is sufficient in order to establish the validity of the RG equation, as can be seen from the analysis of Sect.~\ref{genth}.

One therefore sees that the RG equation is again compatible with the WPC in the physically relevant gauge-invariant sector of the theory.

\section{\label{concl}Conclusions and outlook}

The existence of a RG equation for the \nlsm and the NLSM, compatible with the WPC, shows that the sliding of the scale $\mu$ on physical amplitudes can be reabsorbed by suitable finite counterterms, arising at the loop order prescribed by the WPC itself. As a result, the running with energy of physical observables becomes a consistent procedure also within non-linearly realized theories based on the LFE and for which the WPC holds.

Moreover one can formulate the notion of weak stability, in close analogy with the power-counting renormalizable case: the inclusion of free finite counterterms at higher order in the loop expansion alters the Hopf algebra of the model and moreover generates $\lambda$-vertices, thus mixing up the order of the counterterms needed to recover the effect of a change in the scale $\mu$ of the radiative corrections. This destroys the compatibility between the WPC and the RG equation.

It is a rather remarkable fact that in the NLSM the RG equation can be written for a theory with a {\em finite} number of  parameters dictated by the WPC.
Consequently, since the RG equation allows to run physical observables with energy, the parameterization of bSM physics embedded in the NLSM can be tested on a wide range of energy, from the GeV to the TeV scale.

This provides a consistent theoretical framework for the study of the non-linear St\"uckelberg-like symmetry breaking contribution (and their bSM implications) to the fermion and gauge bosons mass generation mechanism, that will be one of the main goals of the next LHC run.

\acknowledgments

One of us (A. Q.) would like to thank C.~Anastasiou for useful discussions.

\end{document}